\documentclass[english]{article}
\usepackage[T1]{fontenc}
\usepackage[latin9]{inputenc}
\usepackage{geometry}
\usepackage{float}
\usepackage{textcomp}
\usepackage{url}
\usepackage{amstext}
\usepackage{graphicx}
\usepackage{setspace}
\usepackage[authoryear]{natbib}
\makeatletter

\providecommand{\tabularnewline}{\\}

\usepackage{ae,lmodern}
\usepackage{lineno}
\usepackage{authblk}
\usepackage[colorlinks=true, allcolors=blue]{hyperref}

\date{}

\author[1,2]{Frédéric Barraquand}
\author[1]{John-André Henden}
\author[3,4]{Olivier Gilg}
\author[1]{\\ Rolf A. Ims}
\author[1]{Nigel G. Yoccoz}

\affil[1]{\normalsize Department of Arctic and Marine Biology, UiT The Arctic University of Norway, Tromsø, Norway}
\affil[2]{\normalsize Institute of Mathematics of Bordeaux, CNRS and University of Bordeaux, Talence, France}
\affil[3]{\normalsize Laboratoire Chrono-environnement, Université de Bourgogne Franche-Comté, Besançon, France}
\affil[4]{\normalsize Groupe de Recherche en Ecologie Arctique, Francheville, France}

\makeatother

\usepackage{babel}
\begin{document}
\title{The Traill island model for lemming dynamics, how it compares to Fennoscandian
vole dynamics models, and a proposed simplification}
\maketitle
\begin{abstract}
The Traill island model of Gilg et al. (2003) is a landmark attempt
at mechanistic modelling of the cyclic population dynamics of rodents,
focusing on a high Arctic community. It models the dynamics of one
prey, the collared lemming, and four predators : the stoat, the Arctic
fox, the long-tailed skua and the snowy owl. In the present short
note, we first summarize how the model works in light of theory on
seasonally forced predator-prey systems, with a focus on the temporal
dynamics of predation rates. We show notably how the impact of generalist
predation, which is able here to initiate population declines, differs
slightly from that of generalist predation in other mechanistic models
of rodent-mustelid interactions such as Turchin \& Hanski (1997).
We then provide a low-dimensional approximation with a single generalist
predator compartment that mimics the essential features of the Traill
island model: cycle periodicity, amplitude, shape, as well as generalist-induced
declines. This simpler model should be broadly applicable to model
other lemming populations that predominantly grow under the snow during
the winter period. Matlab computer codes for Gilg et al. (2003), its
two-dimensional approximation, as well as alternative lemming population
dynamics models are provided. 
\end{abstract}

\section*{The Gilg, Hanski \& Sittler (2003) model for Arctic lemmings}

The predator-prey community in \citet{gilg2003cyclic} is constituted
of one prey species, the collared lemming (\emph{Dicrostonyx groenlandicus}),
and its four predators: the stoat (\emph{Mustela erminea}), the Arctic
fox (\emph{Alopex lagopus}), the long-tailed skua (\emph{Stercorarius
longicaudus}) and the snowy owl (\emph{Bubo scandiacus}). The basic
structure of the model is that of a coupled system of nonlinear differential
equations for the lemming and stoat populations (present year-round
and all years), with time-varying terms. The time-varying part of
the model results largely from avian and fox predation on lemmings,
which happens only in the summer, in addition to a heightened intrinsic
population growth of lemmings in winter. Finally, another forcing
term comes from the stoat reproduction, which is modelled as a discontinuous
burst, the stoat population being multiplied by $(1+v)$ every year
in the spring. The model here is slightly reformulated to make its
mathematical structure more apparent. The lemming population dynamics
are described by

\begin{equation}
\frac{dN}{dt}=\underbrace{r(t)N}_{\text{exp. growth}}-\underbrace{\Gamma(N,N',t)}_{\text{generalist predation}}-\underbrace{\frac{cN^{2}P}{D^{2}+N^{2}}}_{\text{specialist predation}}.
\end{equation}

For convenience, we will count time in unit of years, and define the
variable $t_{\text{mod}}=t\equiv1$, hence $t_{\text{mod}}$ is time
of year between 0 and 1. A key variable is $N'$, the lemming density
at snowmelt: 

\begin{eqnarray}
\begin{array}{cc}
t_{\text{mod}}<t_{\text{snowmelt},} & N'=N(t)\,\&\,r(t)=r_{w}\\
t_{\text{mod}}>t_{\text{snowmelt}}, & N'=N(t_{\text{snowmelt}})\,\&\,r(t)=r_{s}
\end{array}
\end{eqnarray}

$N'$ can be thought of as a perceived lemming density by generalist
and nomadic predators upon their seasonal arrival to the system, that
introduces a short time delay in summer in the model (decisions made
by the predators are conditional to $N'$). The generalist predation
term $\Gamma(N,N',t)$ is exactly zero in winter, and changes during
the summer as a function of settlement and reproduction schedules
of the various predators. The stoat density $P$ has dynamics of the
form 

\begin{equation}
\frac{dP}{dt}=-(d_{h}+\Delta(N)(d_{l}-d_{h}))P
\end{equation}

with $d_{h}$ the maximum stoat death rate and $d_{l}$ the minimum
death rate, and $\Delta(N)$ a sigmoid function between 0 and 1, that
makes the dynamics switch between the two mortality rates according
to the formula $\Delta(N)=1/2+\arctan(b(N-D))/\pi$. In other words,
there is a higher predator death rate when there is no food. The stoat
compartment is additionally subjected to an interruption and modification
of the state variable, i.e., each year at time $t_{\text{stoat}}$
the integration stops and the predator density switches from $P$
to $P(1+v)$ where $v$ is the number of offsprings (the youngs are
assumed to be equivalents to adults). 

The generalist predation rate can be decomposed into 3 separate terms
corresponding to the different predators (both for adults and juveniles,
the latter being counted in ``adult equivalents''). It is a function
of time through the predator densities: 

\begin{equation}
\Gamma(N,N',t)=\underbrace{\frac{W_{f}N^{2}(P_{f}(t)+P_{yf}(t))}{D_{f}^{2}+N(t)^{2}}}_{\text{fox}}+\underbrace{\frac{W_{o}N(t)^{2}(P_{o}(t)+P_{yo}(t))}{D_{o}^{2}+N(t)^{2}}}_{\text{owl}}+\underbrace{\frac{W_{l}N(t)^{4}(P_{l}(t)+P_{yl}(t))}{D_{l}^{4}+N(t)^{4}}}_{\text{skua}}.
\end{equation}

The numerical response of the predators (and hence, the seasonal variation
in generalist predation pressure) is fully described in Table 1. 

\begin{table}[H]
\begin{raggedright}
{\footnotesize{}}%
\begin{tabular}{|c|ccccc|}
\hline 
{\footnotesize{}Predator} & {\footnotesize{}Adults (when present)} & {\footnotesize{}Youngs x Growth youngs} & {\footnotesize{}Arrival date} & {\footnotesize{}Leaving date} & {\footnotesize{}Birth date}\tabularnewline
\hline 
{\footnotesize{}Fox} & {\footnotesize{}$P_{f}=\frac{b_{f}N'^{2}}{Y_{f}^{2}+N'^{2}}$} & {\footnotesize{}$P_{yf}(t)=\frac{b'_{f}N'^{2}}{Y_{f}\text{'\texttwosuperior}+N'^{2}}$
$\times$$\frac{1}{1+e^{-0.36(365t_{\text{mod}}-9)}}$} & {\footnotesize{}$t_{ofa}=0.52$} & {\footnotesize{}$t_{\text{fall}}=1.0$} & {\footnotesize{}$t_{\text{snowmelt}}=0.65$}\tabularnewline
{\footnotesize{}Owl} & \multicolumn{1}{c}{{\footnotesize{}$P_{o}=\frac{b_{o}(N'-2)}{Yo+N'-4}$}} & {\footnotesize{}$P_{yo}(t)=\frac{b'_{o}(N'-2)}{Y'_{o}+N'-4}$$\times$$\frac{1}{1+e^{-0.36(365t_{\text{mod}}-9)}}$} & {\footnotesize{}$t_{ofa}=0.52$} & {\footnotesize{}$t_{ol}=0.94$} & {\footnotesize{}$t_{\text{birth owl}}=0.67$}\tabularnewline
{\footnotesize{}LT skua} & {\footnotesize{}$P_{l}=0.02$} & {\footnotesize{}$P_{yl}(t)=\frac{b'_{l}N'^{2}}{Y_{l}'^{2}+N'^{2}}$$\times$$\frac{1}{1+e^{-0.464(365t_{\text{mod}}-4.55)}}$} & {\footnotesize{}$t_{la}=0.62$} & {\footnotesize{}$t_{ll}=0.81$} & {\footnotesize{}$t_{\text{birth skua}}=0.72$}\tabularnewline
\hline 
\end{tabular}{\footnotesize\par}
\par\end{raggedright}
\caption{Numerical responses of generalists. The density of adults in the first
column apply only during the period between the arrival and leaving
dates mentioned in the 4th and 5th columns. The density of youngs
is conditional on that of the adults being positive, and will be non-zero
after the birth date. $P_{o}=0$ whenever $N'<2$. Note, for comparison,
that the time of stoat reproduction is $t_{\text{stoat}}=0.69$. $t_{ofa}=$
arrival time for the owl and fox, $t_{la}=$ arrival time for the
long-tailed skua, $t_{ol}=$ leaving time for the owl, $t_{ll}$=
leaving time for the long-tailed skua.}
\end{table}

\begin{figure}[H]
\begin{centering}
\includegraphics[scale=0.9]{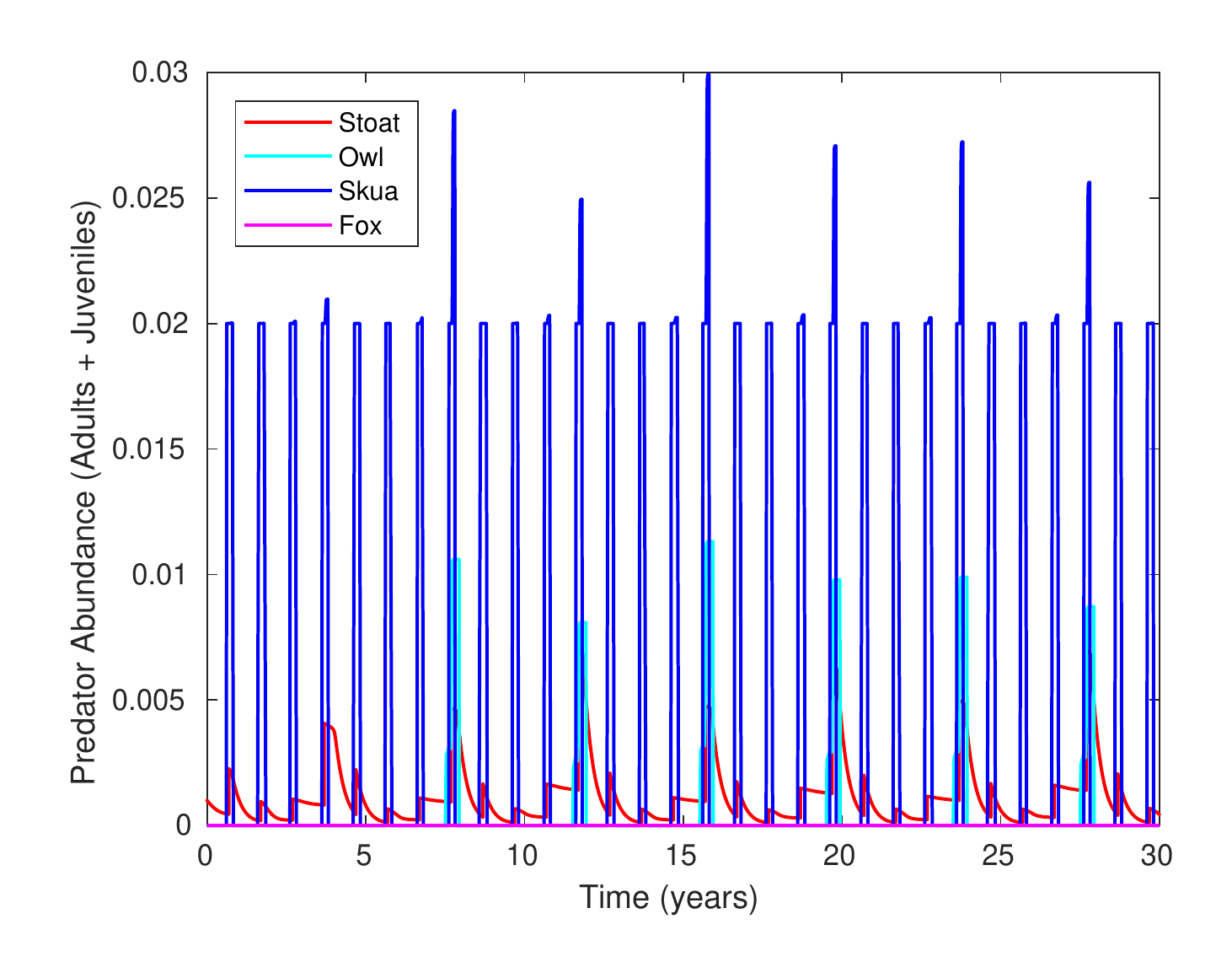}
\par\end{centering}
\caption{\textbf{Predator abundances} over time in the Gilg et al. (2003) model
for the reference parameter set with all predators present.}
\end{figure}

\begin{figure}
\begin{centering}
\includegraphics[width=15cm]{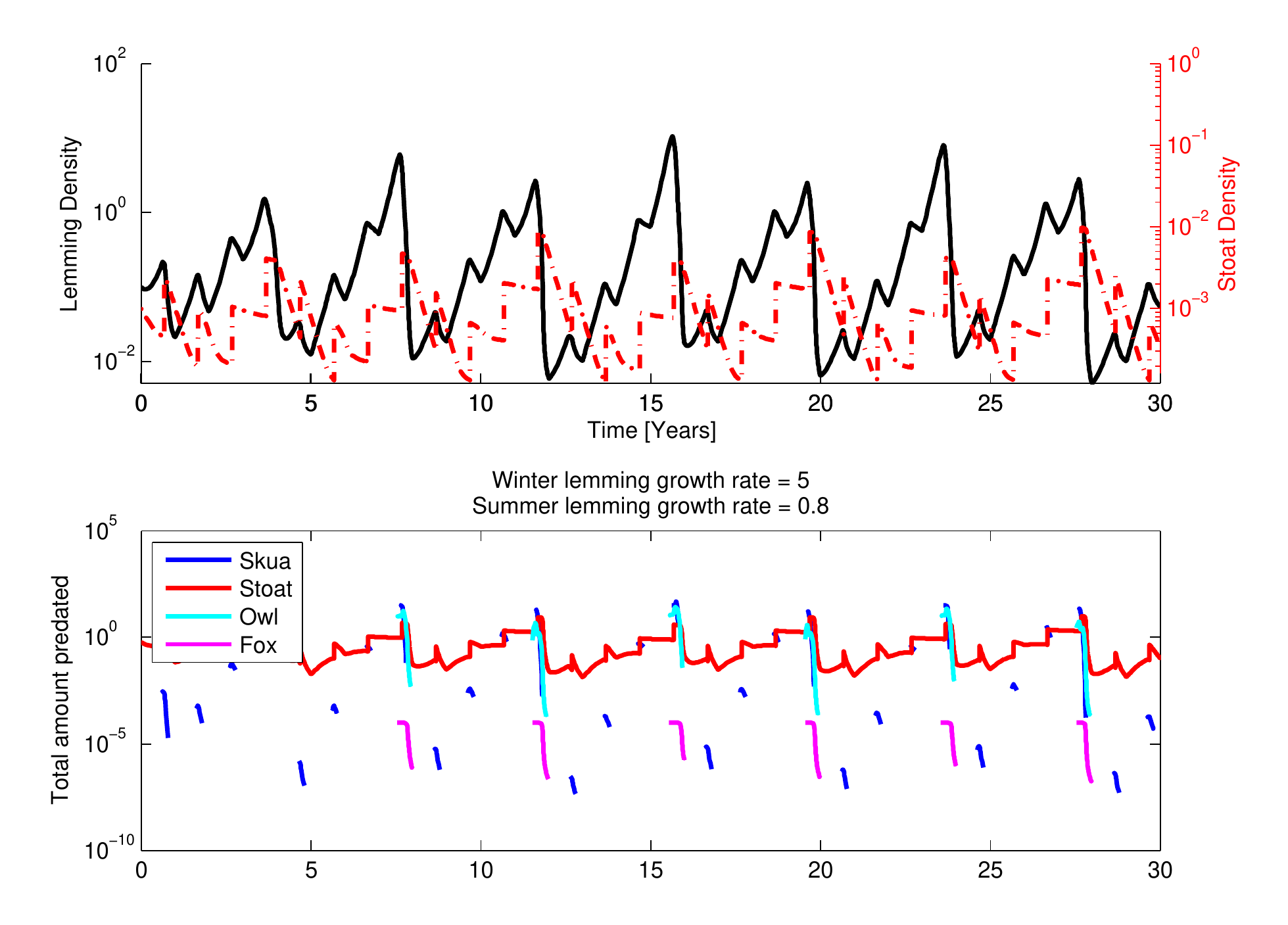}
\par\end{centering}
\caption{\textbf{Population cycles and predation rates }in the Gilg et al.
(2003) model, in logarithmic scale. Parameters for lemming and stoats:
$r_{W}=5,\,r_{S}=0.8,\,v=4.0,\,c=1000,\,D=0.08,\,N_{crit}=D,\,d_{l}=0.1,\,d_{h}=4,\,b=25$.
Lemming density is given in individuals per ha. }
\end{figure}

This seasonal and large mortality ($\approx$ 80 to 90\% of lemming
individuals are eaten by skuas and owls over the summer in peak years\footnote{computed for a few peaks with the model})
is in effect \emph{equivalent to a very large seasonal perturbation},
mirroring theoretical results that show the oscillation-generating
effects of such seasonal perturbations \citep{rinaldi1993multiple,king2001gpc,taylor2012seasonal}.
Simulations of another, simplified Lemming-Stoat-Skua (LSS) model\footnote{which assumes that all generalist predators behave like skuas}
adapting the framework of \citet{turchin1997ebm} to reduce the model
complexity of the Gilg et al. (2003) model, show that the 95\% upper
quantile of lemming values can be increased by a factor of about 1.3
in case of seasonal rather than constant generalist predation. Hence
seasonal generalist predation, together with other sources of seasonality
(e.g., in birth rates, \citealp{taylor2013variations}), can increase
the potential for high-amplitude oscillations. Our LSS model does
confirm, however, that increases in the \emph{average} quantity of
generalists (\emph{G}) such as skuas decreases cycle amplitude and
periodicity like shown in \citet{turchin1997ebm}. 

The \citet{gilg2003cyclic} model without mustelids, but with generalist
predators, can exhibit 2-year population cycles for some parameter
values (Fig. 3), and\textbf{ }this is largely due to the recruitment
of juveniles foxes at the end of the year. We initially spotted this
because of a typo in \citet{gilg2003cyclic}'s Supplementary Material
(which has been corrected in \citealp{gilg2009climate} and did not
affect \citealp{gilg2003cyclic}'s simulations) where the max density
of owls $b_{0}$ had been multiplied by two. Two-year population cycles
do not appear for the standard parameter set of \citet{gilg2003cyclic},
but it is easy to imagine that for a slightly different predator composition
at another study site, such short-term fluctuations might become possible. 

\begin{figure}
\begin{centering}
\includegraphics[width=14cm]{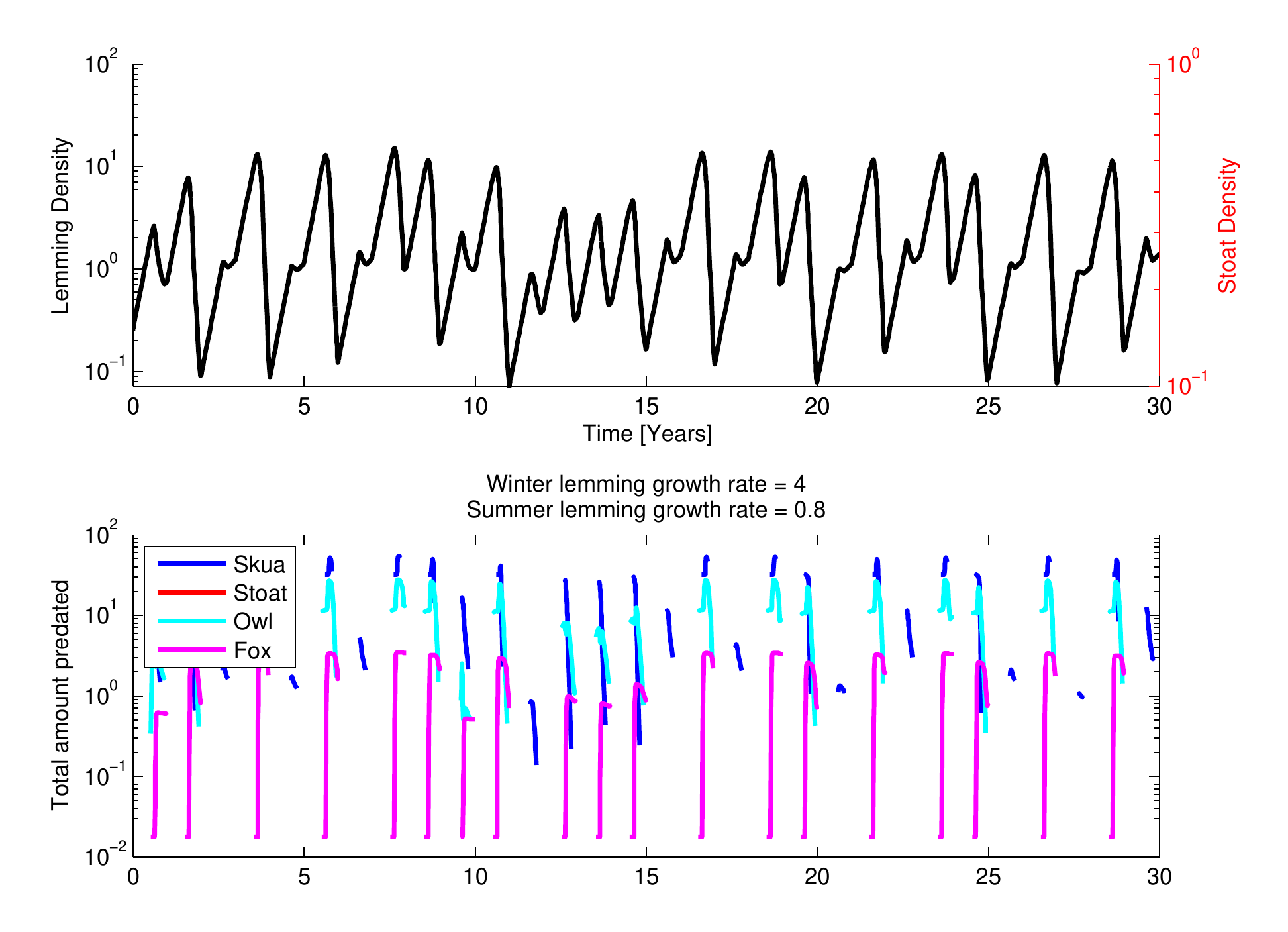}
\par\end{centering}
\caption{\textbf{2-year population cycles (and transients) in the Gilg et al.
(2003) model without mustelids. }Parameters for lemming growth: $r_{W}=4,\,r_{S}=0.8,\,v=4.0,\,c=1000,\,D=0.1,\,N_{crit}=D,\,d_{l}=0.1,\,d_{h}=4,\,b=25$.
The maximum number of owls has been multiplied by two. }
\end{figure}

On the other hand, removing the skua population in the differential
equations model leads to a blow-up (unstable oscillations diverging
away from the attractor). Therefore, our interpretation of the cycles
in the model of \citet{gilg2003cyclic} is that: 
\begin{enumerate}
\item Long-tailed skuas keep the lemming cycle within bounds through predation
during the summer and generate, together with owls, strong seasonal
forcing. Their influence is therefore two-fold: (a) their generalist
predation tends to keep lemmings in check but (b) the fact that such
predation is seasonal contributes to the population cycling. 
\item The slightly delayed reproductive response of foxes can for some parameter
sets create a short-period (2-year) cycle in absence of mustelids. 
\item Stoats generate an eventually unstable lemming-stoat oscillation,
which is transformed into a more sustainable attractor by generalists
(skuas in particular). 
\end{enumerate}
Our conclusion is therefore that both `generalists' such as skuas/owls/foxes
and specialists such as mustelids contribute to some degree to generating
collared lemming fluctuations in the Traill Island model. These considerations
open up new challenges in defining the precise role of generalist
(or nomadic specialists) versus resident specialist predators. 

\section*{Comparison to the vole-weasel model of Turchin and Hanski (1997)}

The model of \citet{turchin1997ebm} can be written

\begin{equation}
\frac{dN}{dt}=rN\left(1-\frac{N}{K}\right)-\underbrace{\frac{GN^{2}}{C^{2}+N^{2}}}_{\text{generalists}}-\underbrace{\frac{aNP}{D+N}}_{\text{specialists}}
\end{equation}

\begin{equation}
\frac{dP}{dt}=sP\left(1-q\frac{P}{N}\right)
\end{equation}

Adding seasonality and adimensionalizing, we arrive at 

\begin{equation}
\frac{dn}{dt}=r(1-e\sin(2\pi t))n-rn^{2}-\frac{gn^{2}}{h^{2}+n^{2}}-\frac{an}{n+d}
\end{equation}

\begin{equation}
\frac{dp}{dt}=s(1-e\sin(2\pi t))p-sp^{2}/n
\end{equation}

with possibly a small noise term on all parameters, so that each parameter
$\Pi_{t}$ is transformed once a year into $\Pi_{t}(1+\sigma\epsilon_{t}),\epsilon_{t}\sim\mathcal{N}(0,1)$.
This models reproduces the Fennoscandian gradient when $G$ is increased
from South to North. We use here the parameters $r=6,\,e=1.0,\,K=150.0,\,s=1.25,\,C=600.0,D=6.0,Q=40.0,\,G=60.0,\,H=15,\,\sigma=0$.
See \citet{taylor2013variations} for a recent investigation of the
possible effects of changes in seasonality over the gradient on cycle
periodicity and amplitude. 

In Fig. 4 below we illustrate the time series of weasel and vole densities
as well as the total amount killed per unit time for specialists or
generalists, which shows that specialist predation is larger than
generalist predation during vole population declines. 

\begin{figure}[H]
\begin{centering}
\includegraphics[width=14cm]{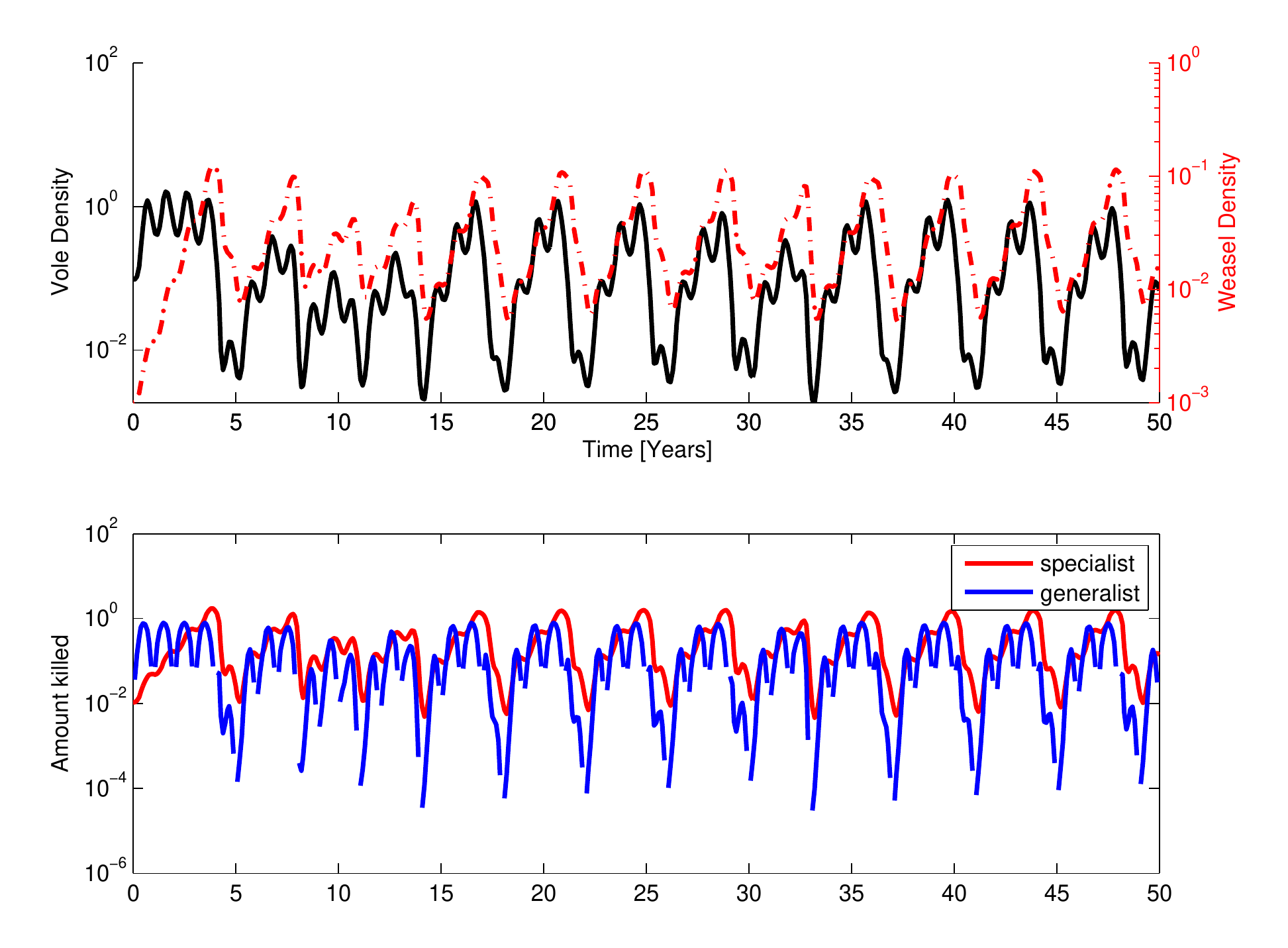}
\par\end{centering}
\caption{\textbf{Densities and predation rates in a modified version of the
Turchin \& Hanski (1997) model}, where we additionally allowed generalist
predation rates to vary seasonally. Voles still reproduce better in
summer than winter, which counteracts the increased predation levels
in summer.\label{fig:modifiedTH97}}
\end{figure}

By contrast, the model that we adapted for modelling Greenland lemmings
using the same model framework shows a slightly different kind of
dynamics. 

\section*{``Pooled generalists'' model (aka Lemming - Stoat - Skua model)}

Because there are very many parameters in \citet{gilg2003cyclic}
(26 in Table S1 and S2 from the original paper, plus potentially 10
dates from Table 1 of this article), we constructed a simplified model
to compare its behaviour to \citet{turchin1997ebm}. The model assumes
that all generalists behave like the skua (the more abundant generalist,
with numbers that only depend on the season and not on other species
densities). It also assumes that the numerical response has the more
simplifed Leslie-type form of \citet{turchin1997ebm}, which allows
to formulate a smooth model. The LSS model uses a winter indicator
variable $W(t)=\frac{1}{2}(1+\cos(2\pi t))$. The time $t=0$ is in
january, so that $W=1$ in full winter, $0$ in full summer. The full
differential equation model then writes

\begin{equation}
\frac{dN}{dt}=r_{min}N+(r_{max}-r_{min})W(t)N-r_{max}\frac{N^{2}}{K}-\underbrace{\frac{G(1-W(t))N^{4}}{H^{4}+N^{4}}}_{\text{generalist = skua}}-\underbrace{\frac{CN^{2}P}{D^{2}+N^{2}}}_{\text{specialist = stoat}}
\end{equation}

\begin{equation}
\frac{dP}{dt}=sP\left(1-q\frac{P}{N}\right)
\end{equation}

\begin{figure}[H]
\centering{}\includegraphics[width=14cm]{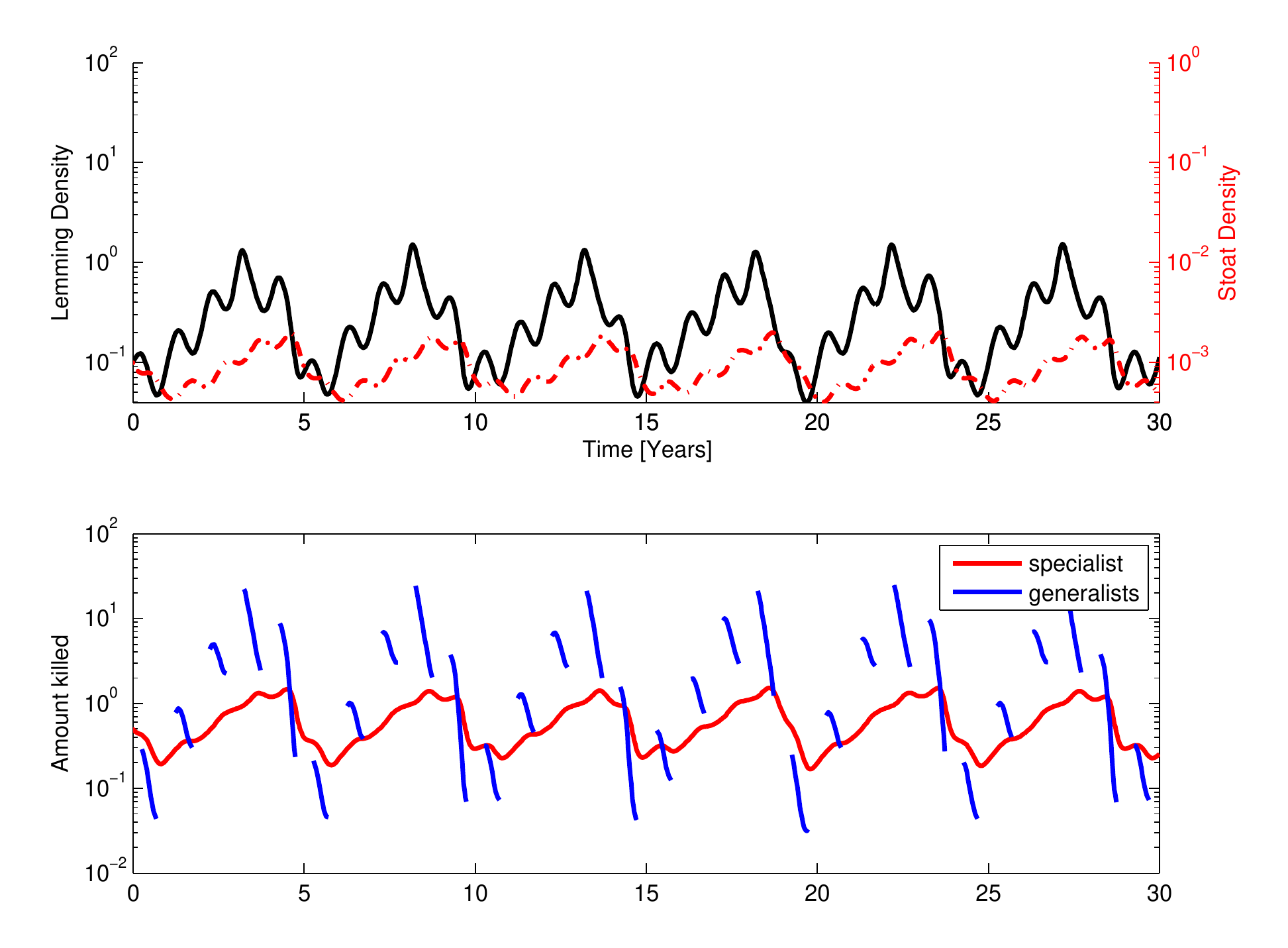}\caption{\textbf{``Pooled generalists'' LSS model simulation}. See text for
parameter values. }
\label{LSS}
\end{figure}

Parameters used in Fig. \ref{LSS} are tailored to the Greenland Traill
island case study: $r_{max}=6,\,r_{min}=0.5,\,K=500,\,G=50,\,H=2,\,C=1000,\,D=0.1,\,s=1.75,\,Q=100.$
Because the carrying capacity $K$ in absence of predation was absent
in the Traill island model, it is here set to a large value, but it
is notable that the LSS model can also work without. The functional
response exponents have been taken in accordance to \citet{gilg2003cyclic}
and are all sigmoid (Type III), in contrast to the more classical
choice of type II response for specialists and type III for generalists
in \citet{turchin1997ebm}. 

Simulating this model, Fig 5 below, we see that the predation by generalists
in summer is much higher than that of the specialist (note the logarithmic
scale), in constrast to \citet{turchin1997ebm}. We also see very
clearly that generalists clearly initiate the lemming declines, by
``cropping off'' the lemming peaks. 

\subsection*{A cautionary tale on cycle shape in mechanistic models}

It has been proposed that cycle shape can be a proxy for cycle causation
\citep{turchin2000lemmings}. We offer a different view here. A difficulty
with the numerous mechanistic mathematical models available to model
rodent cycles is that they tend to produce cycles with correct periodicity
and amplitude, but with shapes often different from that of the data.
In other words, mechanistic models with empirically estimated parameters
might not fit all the details of the cycle very well. For example,
in \citet[Fig. 4]{korpimaki2002dynamic} the mechanistic predation
model consistently produces cycles that rank in the right side of
Royama's triangle, where delayed density-dependence generates the
crashes after a plateau at high density \citep{royama1992analytical}.
However, the data shows often faster crashes, ranking on the left
side of the triangle (Fig. 4 in their paper). Conversely, the model
in \citet{gilg2003cyclic} cited above tends to produce fast crashes
in just one year while the data show crashes occurring over one or
two years. This is arguably an undesirable property of the model \citep{oksanen2008arctic},
which incidentally refutes the claims of \citet{turchin2000lemmings}
that ``prey peaks'' have a rounded shape. The models by \citet{korpimaki2002dynamic}
and \citet{gilg2003cyclic} therefore seem to be equally good at describing
periodicity and amplitude of the time series, but equally limited
to reproduce cycle shape\footnote{This remark may very well apply to numerous other mechanistic models
for cyclic populations.}. With rich datasets (e.g. >100 data points in this context), it might
be possible to find the most likely models just based on cycle shape,
but given the limited data available here ($\approx$ 25 years with
no spatial replication, which is common in many stuch study sites),
such endeavours may be a little premature. Mechanistic mathematical
models may be best interpreted as illustrating what is \emph{possible,}
rather than what is actually happening in real populations. Although
in some cases, it may be possible to at least rank different scenarios
based on very contrasted models and multiple model diagnostics \citep{kendall2005population}. 

\subsubsection*{Code availability}

Computer codes have been deposited at Zenodo as part of our release
of lemming population cycles models \citep{frederic_barraquand_2020_4271834}
with \url{DOI:10.5281/zenodo.4271833}. These are additionally available
at \url{https://github.com/fbarraquand/lemmingCycles_ODEmodels}.
The repository includes of course code for the original Traill island
model, codes to simulate the Turchin \& Hanski (1997) model with/without
seasonal generalist predation, as well as our proposed LSS simplification
of the model with 9 parameters. The repository also includes features
not used here, such as a stochastic (SDE) version of the Traill island
model, various models considered in \citet{turchin2001availability}
for lemming-plant interactions, and a host-parasite model of an interaction
with unknown parasite, whose main purpose is to elicit caution regarding
inferring cycle causation without hard data on all main parameters
of an interaction. 

\subsubsection*{Acknowledgements}

FB thanks Rachel A. Taylor for discussions and sharing results on
the effects of generalist predation in the Turchin \& Hanski (1997)
model. 

\bibliographystyle{amnat}
\bibliography{traill_model}

\end{document}